\documentclass[twocolumn,fleqn,natbib]{svjour2}
\bibpunct{[}{]}{;}{n}{}{,} 
\smartqed  
\usepackage{graphics,graphicx,dcolumn,bm,fleqn,epic,eepic,float}
\usepackage{amssymb,amsmath,multirow,rotate,color,float,times}
\usepackage{color}
\definecolor{red}{rgb}{1,0,0}
\definecolor{green}{rgb}{0,1,0}
\definecolor{blue}{rgb}{0,0,1}

\journalname{Granular Matter}
\begin{document}

\title{Avalanches in anisotropic sheared granular media}

\author{Andr\'es A.~Pe\~na    \and
        Sean McNamara         \and
        Pedro G.~Lind         \and
        Hans J.~Herrmann}

\authorrunning{A.A.~Pe\~na et al}

\institute{A.A.~Pe\~na, S.~McNamara, P.G.~Lind \at
           Institute for Computational Physics, 
           Universit\"at Stuttgart, Pfaffenwaldring 27, 
           D-70569 Stuttgart, Germany \\
           \email{andres@icp.uni-stuttgart.de}
           \and
           H.J.~Herrmann    \at
           Departamento de F\'{\i}sica, Universidade Federal do Cear\'a,
           60451-970 Fortaleza, Cear\'a, Brazil;\\
           Computational Physics, IfB, HIF E12, 
           ETH H\"onggerberg, CH-8093 Z\"urich,
           Switzerland}

\date{Received: date }

\maketitle

\begin{abstract}
We study the influence of particle shape a\-ni\-so\-tro\-py on the 
occurrence of avalanches in sheared granular media. We use molecular 
dynamic simulations to calculate the relative movement of two tectonic 
plates. 
Our model considers irregular polygonal particles constituting  
the material within the shear zone. We find that the magnitude of 
the avalanches is approximately independent on particle shape and 
in good agreement with the Guten\-berg-Richter law, but the aftershock 
sequences \ are stron\-gly influenced by the particle anisotropy yielding 
variations on the exponent characterizing the empirical Omori's law. 
Our findings enable one to identify the presence of anisotropic
particles at the macro-mechanical level only by observing the avalanche
sequences of real faults. In addition, we calculate the probability of 
occurrence of an avalanche for given values of stiffness or frictional 
strength and observe also a significant influence of the particle 
an\-iso\-tro\-py.
\end{abstract}



\section{Introduction}
\label{sec:intro}

Natural earthquakes are one of the most catastrophic events in 
nature \cite{bolt05} with deep social implications, in terms 
of human casualties and economic loss.
Considerable efforts have been made to understand the earthquake dynamics 
and the underlying mechanisms prior to their occurrence 
\cite{sykes99,donze94,mora02,ramos07}, 
either through experimental studies \cite{byerlee77,marone98,marone02} or
through particle based numerical models
\cite{mora94,tillemans95,mora99,alonso06}
of the relative motion of tectonic plates \cite{bolt05,scholz02,turcotte02}. 

However, in most of the existing numerical models of earthquake 
fault the gouge is represented by discs \cite{mora99,alonso06} 
or spheres \cite{mora06}. 
The dynamics of such material within the fault is thought to control the 
stick-slip instability that characterizes earthquake process. An 
understanding of its properties is, therefore, vital to understand earthquake 
dynamics \cite{wilson05}. For instance, the existence of the gouge within 
the fault has been proposed to explain the low dissipation on shear zones,
providing new insight into the heat flow paradox \cite{mora98}. 
In this case, the reduction of the macroscopic friction and consequently, 
the heat generation is attributed to the deformational patterns such as 
rolling of particles \cite{mora99,alonso06}. In laboratory experiments by 
Maron \cite{marone02}, the influence of particle characteristics has also 
been studied. They found that frictional strength and stability of the 
granular shear zone is influenced by particle shape, size distribution 
and their evolution through particle crushing. 
Therefore, to model fault gouges one must also include different grain 
characteristics.

In this paper, we use a model of polygonal particles \cite{pena07,pena07a} 
to address the influence of anisotropy in granular media.
We study the situation of two tectonic plates with boundaries
parallel to the direction along which the tectonic 
plates move -- so-called transform boundaries \cite{bolt05,turcotte02}.
One of the most well known examples 
of such boundaries is the San Andres Fault in California 
where the Pacific plate and the North American plate are moving in 
opposite directions. In this particular case, the relative motion 
of the plates is about 40 mm/year, and thus the strain accumulation 
rate is around  $3\cdot10^{-7}$ per year \cite{kanamori01}. 
Further, different from previous studies \cite{tillemans95}, our model 
considers anisotropic particle shapes. The response of the system is 
characterized by discrete events or avalanches whose size covers many 
orders of magnitude, similar to the so-called crackling noise of  
physical systems \cite{sethna01}.

We find that the magnitude of the avalanches is independent of the
particle shape and in agreement with the Guten\-berg-Richter 
law \cite{gutenberg54}. On the contrary,
the distribution of the waiting times described by the Omori's 
law \cite{omori95}  strongly depends on shape anisotropy.
From this result, we raise the hypothesis of identifying 
at the macro-mechanical level the presence of 
anisotropic particles within the gouge,
only by studying the temporal avalanche sequences.
We further argue that the existence of this anisotropic gouge in fault 
zones might also explain the variation of the decay of the aftershock 
sequences observed in nature.

In addition, we also compute the conditional probability for an avalanche 
to occur, and found that it decreases logarithmically with the stiffness.
This exponential decay also depends on particle shape, since
anisotropic samples are able to mobilize a higher frictional strength when
compared to isotropic samples. 
For a given value of mobilized strength anisotropic samples also exhibit 
lower probability of failure. 
Finally, we propose some microstructural features that could help to explain 
the occurrence of avalanches.

We start in Sec.~\ref{sec:model} by describing in detail our model of irregular
particles as well as the details of our numerical experiment.
In Sec.~\ref{sec:events} we characterize and study the system response.
In Secs.~\ref{sec:gr} and \ref{sec:waiting} we address the influence
of particle anisotropy on the frequency distribution of avalanches
and on the width of the time interval where aftershocks occur.
The weakening and stability of the system is investigated in 
Sec.~\ref{sec:precursors}, and in Sec.~\ref{sec:conclusions}
the main conclusions are discussed.
\begin{figure}[t]
\begin{center}
\includegraphics*[width=8.5cm,angle=0]{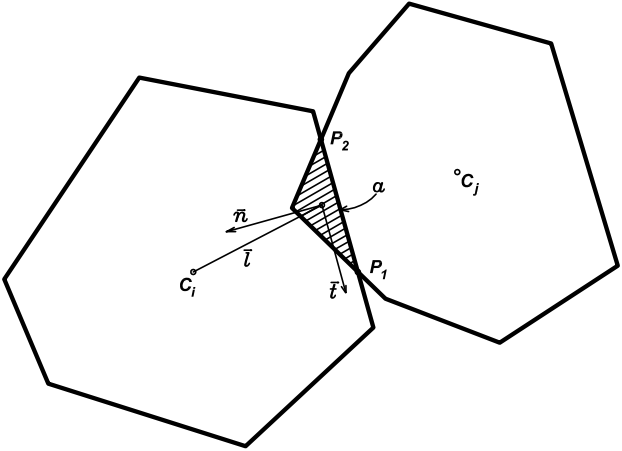}
\end{center}
\caption{\protect
         Schematic representation of a particle contact.
         The overlapping area $a$ is indicated
         by the shaded zone.}
\label{fig1}
\end{figure}

\section{The model}
\label{sec:model}

We consider a two dimensional model of convex polygons to represent 
the grains of the granular material on a mesoscopic scale. 
The samples consist of isotropic and
ani\-so\-tropic particles in order to study the influence of particle shape 
ani\-so\-tropy on the response of a granular packing under very slow shear.

The deformation of the grains is modeled by letting them overlap,
as sketched in Fig.~\ref{fig1}.
When two polygons overlap, the intersection points between their edges can 
be defined. The segment that connects these points, $P_1$ and $P_2$, 
gives the contact line $\vec{S}$ = $P_1 P_2$. The contact 
force is given by $\vec{f}^c = \vec{f}^e + \vec{f}^v$, where $\vec{f}^e$ 
and $\vec{f}^v$ are the elastic and viscous contribution. 

The elastic part of the contact force is decomposed as 
$\vec{f}^e = f^e_n \hat{n}^c + f^e_t \hat{t}^c $, where $\hat{n}^c$ and 
$\hat{t}^c$ are the unitary vectors perpendicular and parallel to the 
contact line $\vec{S}$ respectively. 
The normal elastic force is calculated as $f^e_n = -k_n 
\delta$, where $k_n$ is the normal stiffness, and $\delta$ the deformation 
length defined in terms of the overlapping area $a$ and the length of the 
contact line, $\delta = a / |\vec{S}|$. 
The friction force is given by an elastic force $f^e_t = - k_t \xi$ 
proportional to the elastic displacement $\xi$ at each contact, with $k_t$ 
the tangential stiffness. 
The elastic displacement $\xi$ is updated as $\xi = \xi' + 
\vec{v}_t^c \Delta t$, where $\xi'$ is the previous length of the 
spring, $\Delta t$ is the time step of the molecular dynamic simulation, 
and $\vec{v}_t^c$ the tangential component of the relative velocity 
$\vec{v}^c$ at the contact.
The length of the tangential spring $\xi$ may increase during the
time that the condition $f_e^t < \mu f_e^n$ is satisfied. 
The sliding condition is enforced keeping constant the elastic displacement 
$\xi$ when the Coulomb limit condition $f_t^c = \mu f_n^c$ is reached and 
$\mu$ is the interparticle friction coefficient.

The viscous force is calculated as 
$\vec{f}^v = - m \nu \vec{v}^c$, where 
$m$ is the effective mass of the two particles in contact and $\nu$ the 
damping coefficient. This force takes into account the dissipation at 
the contact and it is necessary to maintain the numerical 
stability of the method.

A suitable closed set of material parameters for this model 
is the ratio $k_t/k_n$, together with the value of 
the normal stiffness $k_n$, the interparticle friction $\mu$, 
and the ratio $\epsilon_t/\epsilon_n$ between the tangential and normal
restitution coefficients.

The random generation of the particles is done by means of a Voronoi 
tessellation using a reference square lattice, yielding a set of nearly 
isotropic polygons. 
By distorting the square lattice in the horizontal and vertical directions,  
we end up with elongated (i.e.~anisotropic) particles. The ratio between the 
stretching and contracting factors gives us the average aspect ratio 
$\lambda$ of the polygons, that is used to characterize the anisotropic 
shape of the particles.

\begin{figure}[b]
\begin{center}
\includegraphics*[width=2.8cm,angle=0]{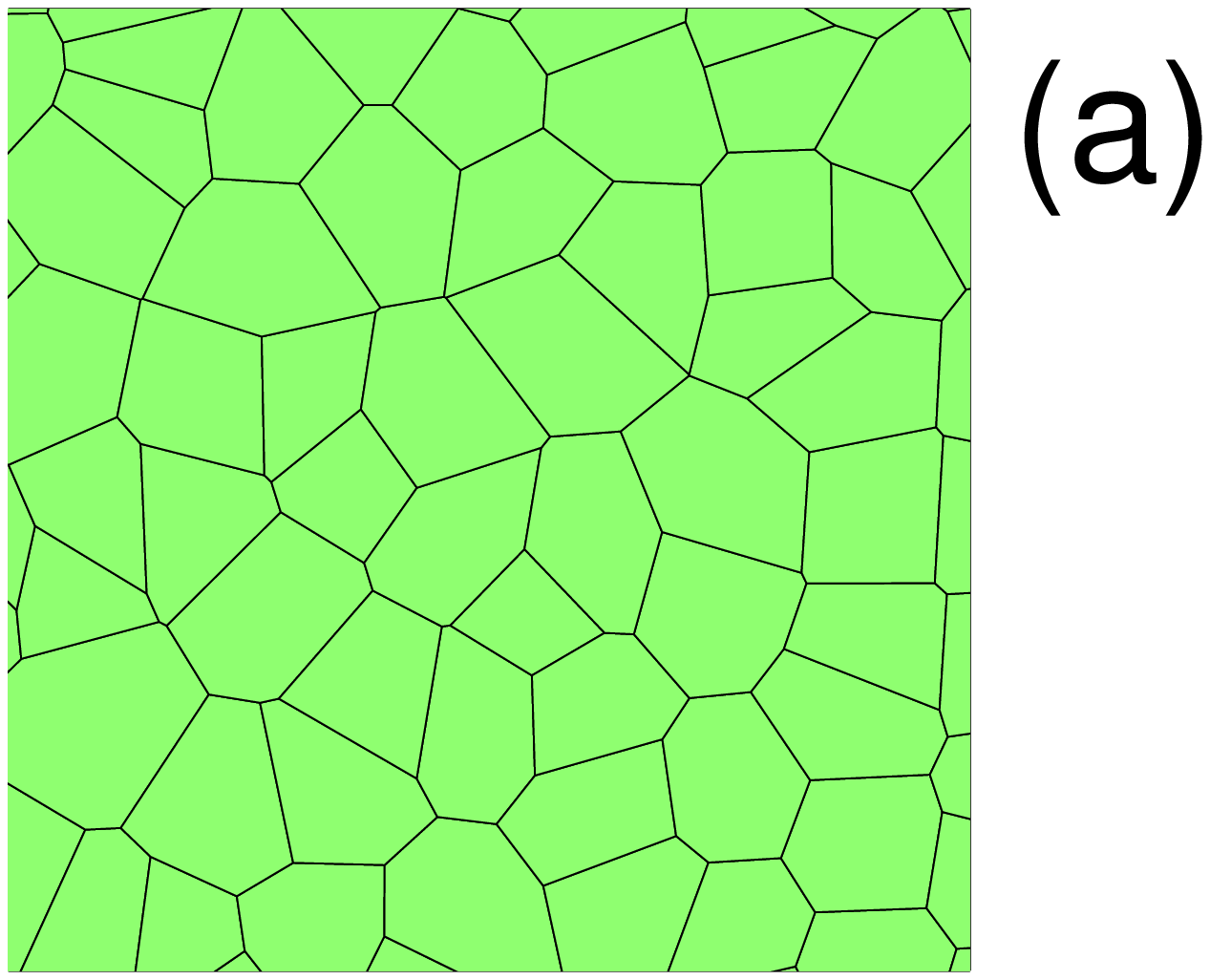}%
\includegraphics*[width=2.8cm,angle=0]{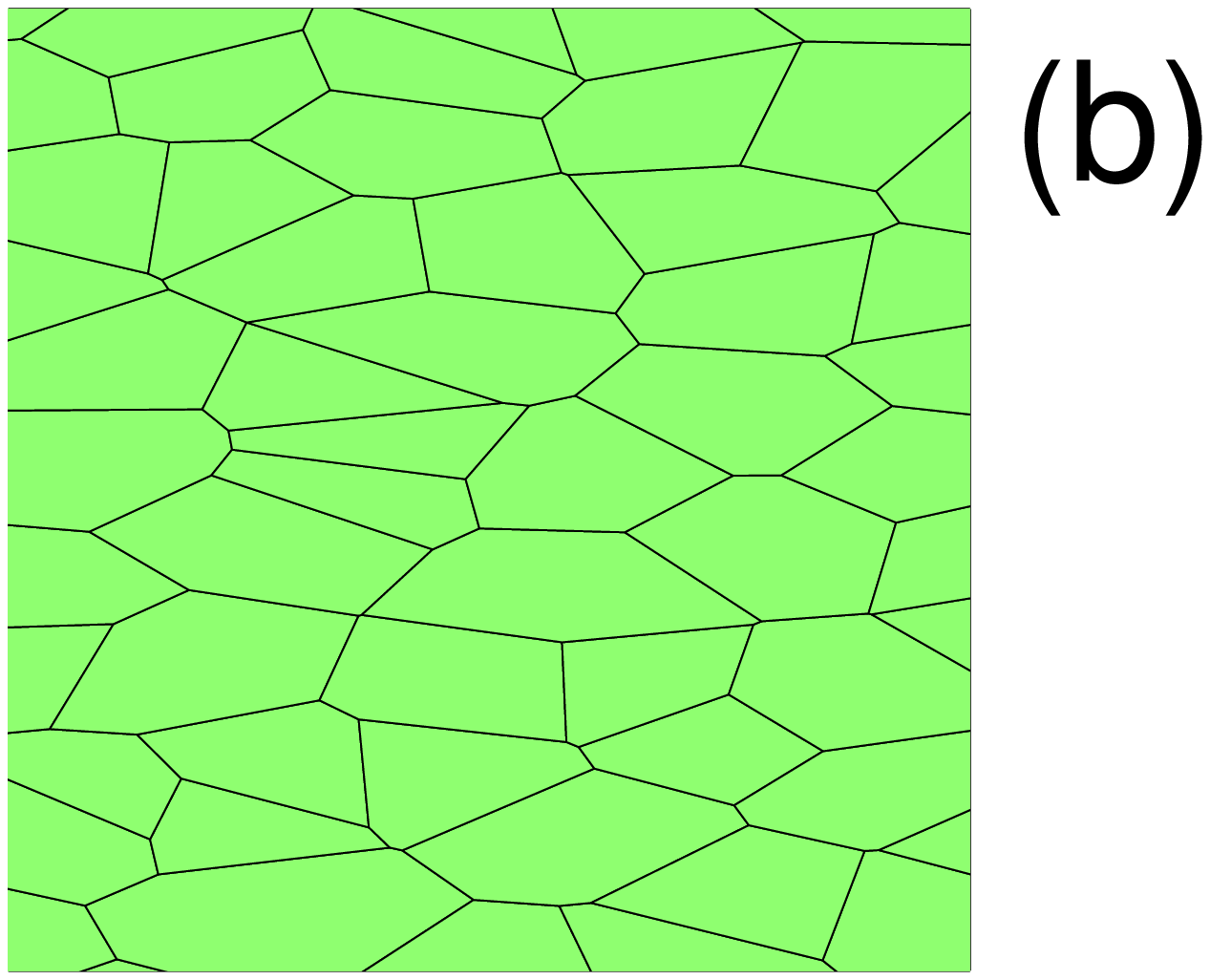}%
\includegraphics*[width=2.8cm,angle=0]{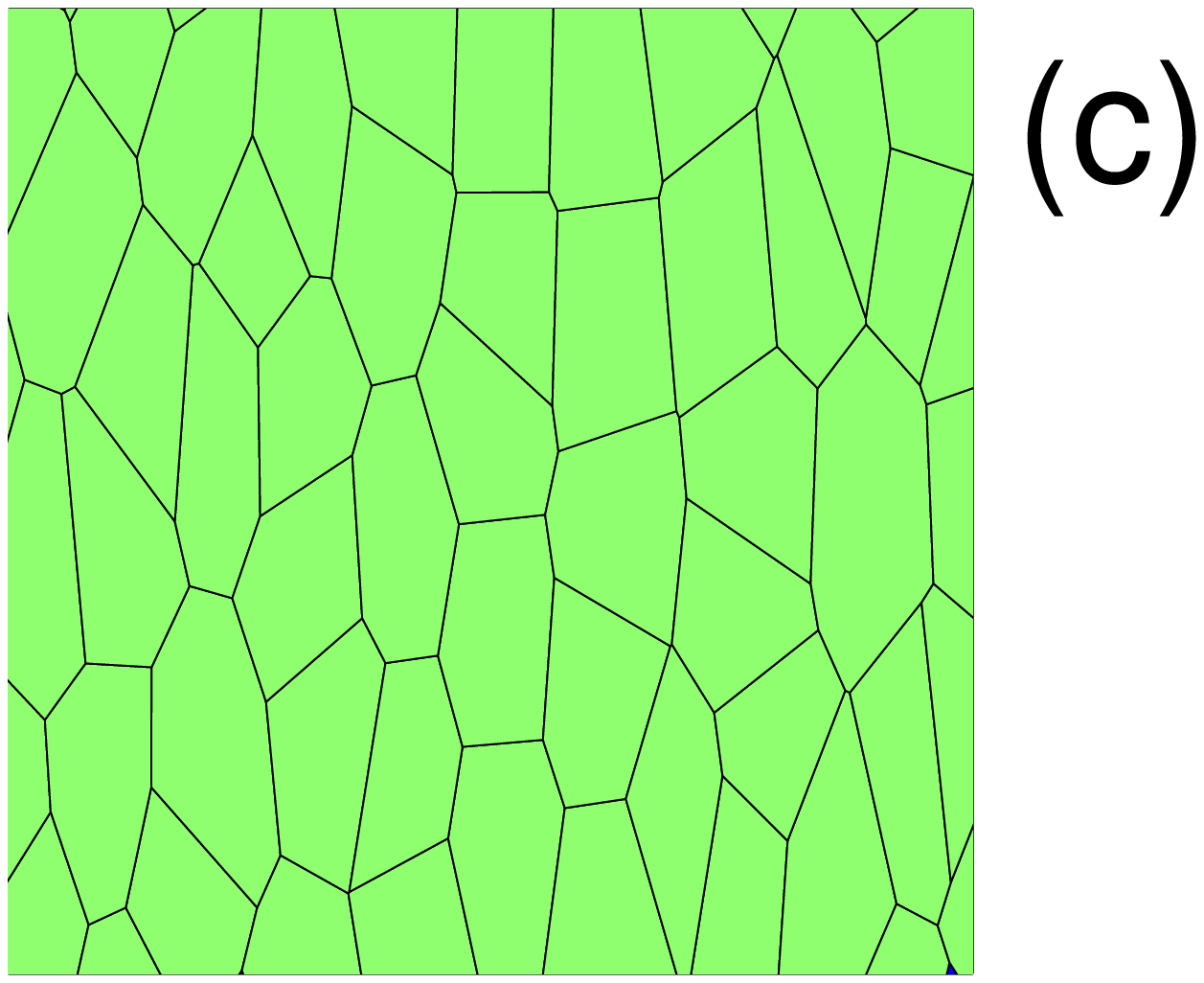}
\end{center}
\caption{\protect
      (Color online) 
      Samples of
      {\bf (a)} isotropic polygons ($\lambda=1$) and 
      {\bf (b)} elongated polygons stretched either in the 
                horizontal direction 
                ($\lambda=2.3$ H) or
      {\bf (c)} in the vertical direction  
                ($\lambda =2.3$ V).}
\label{fig2}
\end{figure}

In Fig.~\ref{fig2} the different initial sample configurations 
are shown. The isotropic configuration is depicted in 
Fig.~\ref{fig2}a, and the anisotropic ones in 
Fig.~\ref{fig2}b for particles stretched in the same direction 
of shearing (horizontal direction, sample H) and in 
Fig.~\ref{fig2}c for particles stretched perpendicularly to the
shear direction (vertical direction, sample V).
\begin{figure}[htb]
\begin{center}
\includegraphics*[width=8.5cm]{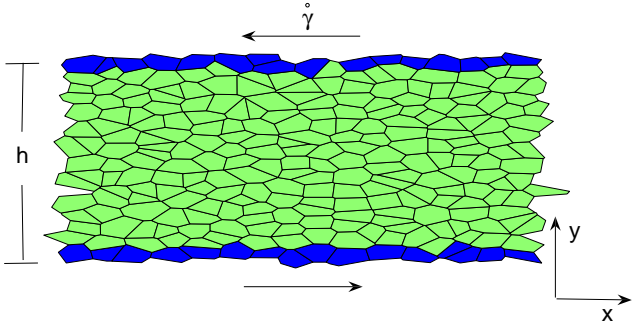}
\end{center}
\caption{\protect
      (Color online) 
         Sketch of the shear cell. 
         The system is not allowed to dilate, i.e.~it has 
         fixed $h$.
         The sample is sheared using a constant shear rate 
         $\dot{\gamma}$. 
         Blue (grey) particles induce shear.}
\label{fig3}
\end{figure}

We use samples of two different sizes, with 256 $(16\times16)$ and 
1024 $(32\times32)$ particles. Periodic boundary conditions are 
imposed in horizontal direction.
A constant horizontal velocity is given to the particles in the top and 
bottom layers so as to impose a constant shear rate $\dot{\gamma}$. 
These particles are not allowed to move in the vertical direction, thus 
suppressing the volumetric strain of the system.
Furthermore, they are not allowed to rotate or move 
against each other. In Fig.~\ref{fig3} 
a setup of the shear cell is presented for the anisotropic sample 
$\lambda = 2.3$ H. The shear strain $\gamma$ is defined as
\begin{equation}
\gamma =  D_x / h,
\end{equation}
where $D_x$ is the horizontal displacement of the boundary
particles and $h$ is the height of the sample, which is kept
constant.

In our simple model, polygons represent rocks composing the gauge
between two tectonic plates and the top and bottom boundary particles 
represent the tectonic plates. 
We start from a perfectly packed configuration in order to represent 
the initial state of the material that is supposed to be intact prior 
to the shear process.
\begin{figure}[htb]
\begin{center}
\includegraphics*[width=9.0cm]{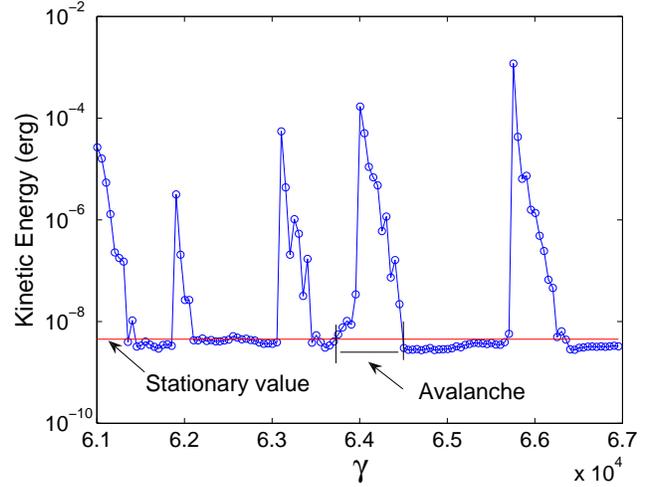}
\end{center}
\caption{\protect
      (Color online) 
        The average kinetic energy in logarithmic scale versus the 
        shear strain $\gamma$. The stationary value $K_0$ of the kinetic
        energy is obtained from the velocity profile of 
        the particles at the steady state. The released energy of the 
        avalanches are calculated using Eq.~(\ref{eq.magnitude}).}
\label{fig4}
\end{figure}

As described above, the value of the strain rate  is of the order of 
$10^{-7}$ per year ($\approx 10^{-14}$ s$^{-1}$). For our numerical 
simulation, this velocity is computationally too expensive. 
For instance, for $\Delta t = 1 s$ one needs $10^{12}$ iterations 
to induce a shear strain of $1 \%$. 
Using a system of 16x16 particles, 
such $10^{12}$ iterations would require roughly $1000$ years of CPU time on 
a standard P-IV PC.
To overcome this, we choose a suitable shear rate at which the motion 
of the system is intermittent, i.e. in some regions the system is locked 
and deforms steadily accumulating elastic strain and in others the 
stored energy at the contacts is suddenly released.
See Fig.~\ref{fig4}.

An important issue in this scope is to study the distribution of the
released energy, spanned over several orders.
We tested shear rates in the range $10^{1} - 10^{-7}$ s$^{-1}$
and found that a suitable value 
for the above purpose is $\dot{\gamma}=1.25\times 10^{-5}$ s$^{-1}$.
Further, to perform the MD simulations using the selected shear rate, 
we adjust the parameters of the model in order to obtain a time step 
$\Delta t$ requiring a reasonable CPU time.
Thus, we use $k_n=400 \hbox{\ N/m}$, $\epsilon_n=0.9875$, 
$k_t/k_n=1/3$, $\nu_t/\nu_n=k_t/k_n$ and $\epsilon_t/\epsilon_n=1.0053$,
yielding a time step of $\Delta t = 0.005 \hbox{\ s}$.
We consider three different interparticle friction coefficients 
$\mu = 0.0, 0.5, 5.0$.
For simplicity we use $\rho=1 \hbox{g/cm}^2$.

\section{System response: monitoring avalanches}
\label{sec:events}

The motion of the particles in the interior of the sample is not continuous, 
but has a ``stick-slip character''. 
During slip a sudden rearrangement of the medium arises as a consequence
of the large relative displacements of the particles. 
We monitor this rearrangement of the system through its kinetic energy  
$K$.
As shown in Fig.~\ref{fig4}, the system can be in two different states.  
In the "steady state", $K$ is approximately equal and less than a low value 
$K_0$, shown by the horizontal line in Fig.~\ref{fig4}.  
This value $K_0$ is associated with the accumulation of elastic strain under 
the imposed shear.  
If the particles were rigid, we would have $K_0=0$, but since we are using 
soft elastic ones, we have $K_0 > 0$.
The low energy state $K_0$ is punctuated by a series of events where kinetic
energy rises several orders of magnitude above $K_0$.  
These are the avalanches.  
An avalanche begins when $K$ rises above $K_0$, and all subsequent values of 
$K$ greater than $K_0$ are considered to be part of the same avalanche.

The total released energy $E_r$ of one avalanche 
is the sum over the total number $N$ of consecutive values of
$K$ above  the stationary state, namely
\begin{equation}
E_r =  \sum_{j=1}^{N} K_j \dot{\gamma}\Delta t , 
\label{eq.magnitude}
\end{equation}
where $\dot{\gamma}\Delta t$ is a proper non-dimensional parameter for integrating
the kinetic energy (see Fig.~\ref{fig4}).
At the `stationary state' the system is deforming steadily and accumulates
energy at the particle contacts. This state can be characterized by the 
value $K_0$ obtained from the average velocity profile of the particles 
at this stage.
\begin{figure*}[!t]
\begin{center}
\includegraphics*[width=7.0cm]{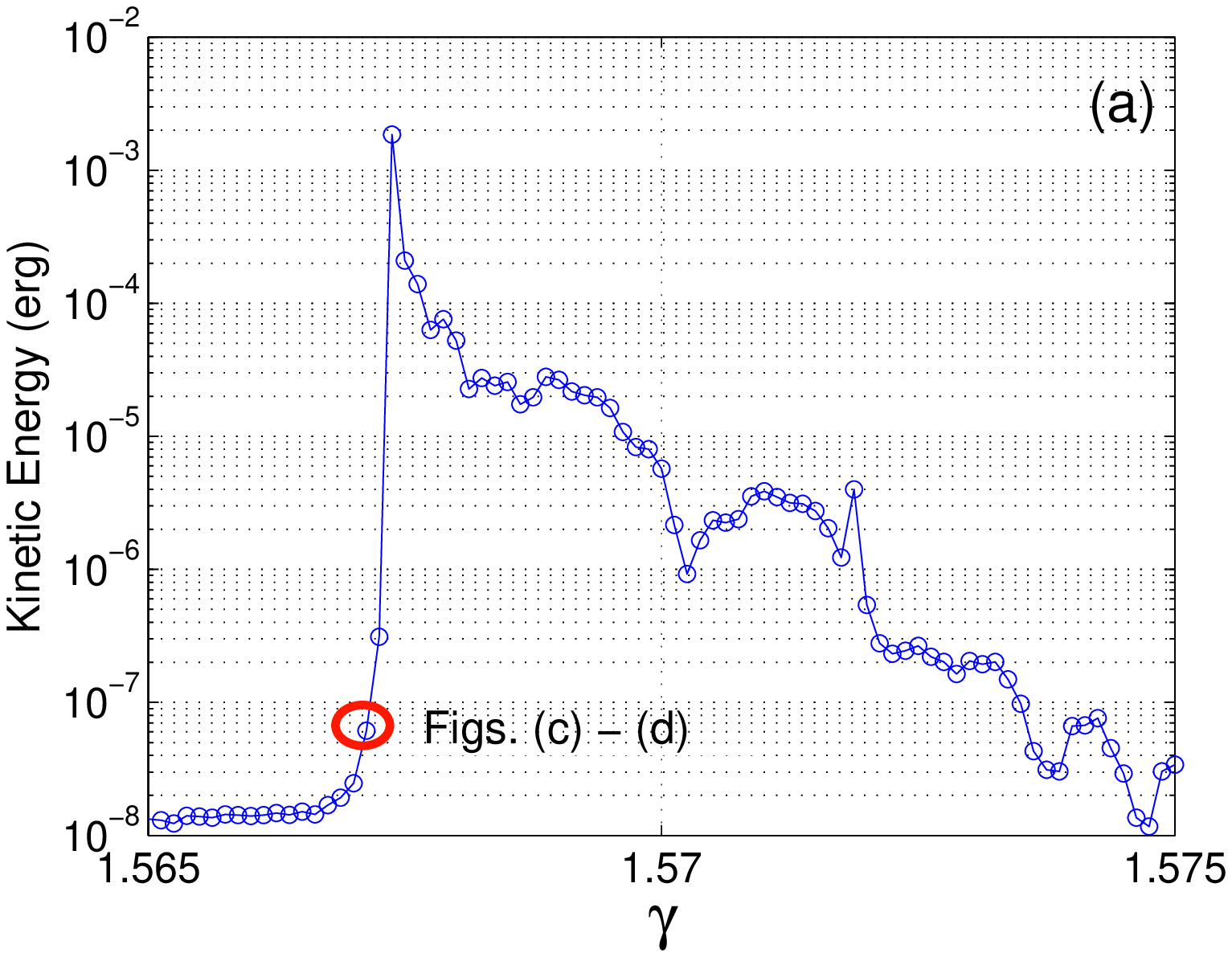}%
\includegraphics*[width=7.0cm]{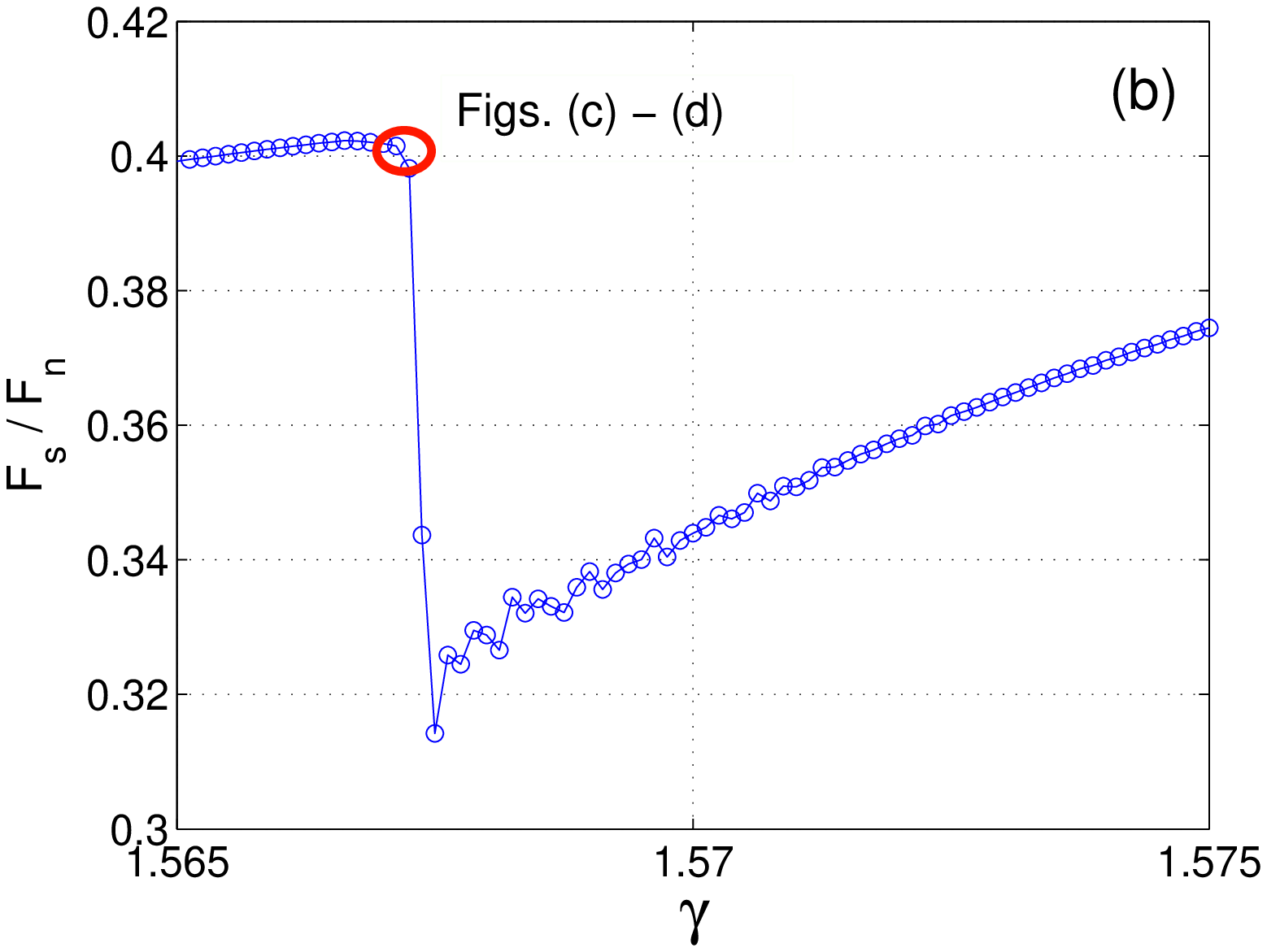}
\includegraphics*[width=7.0cm]{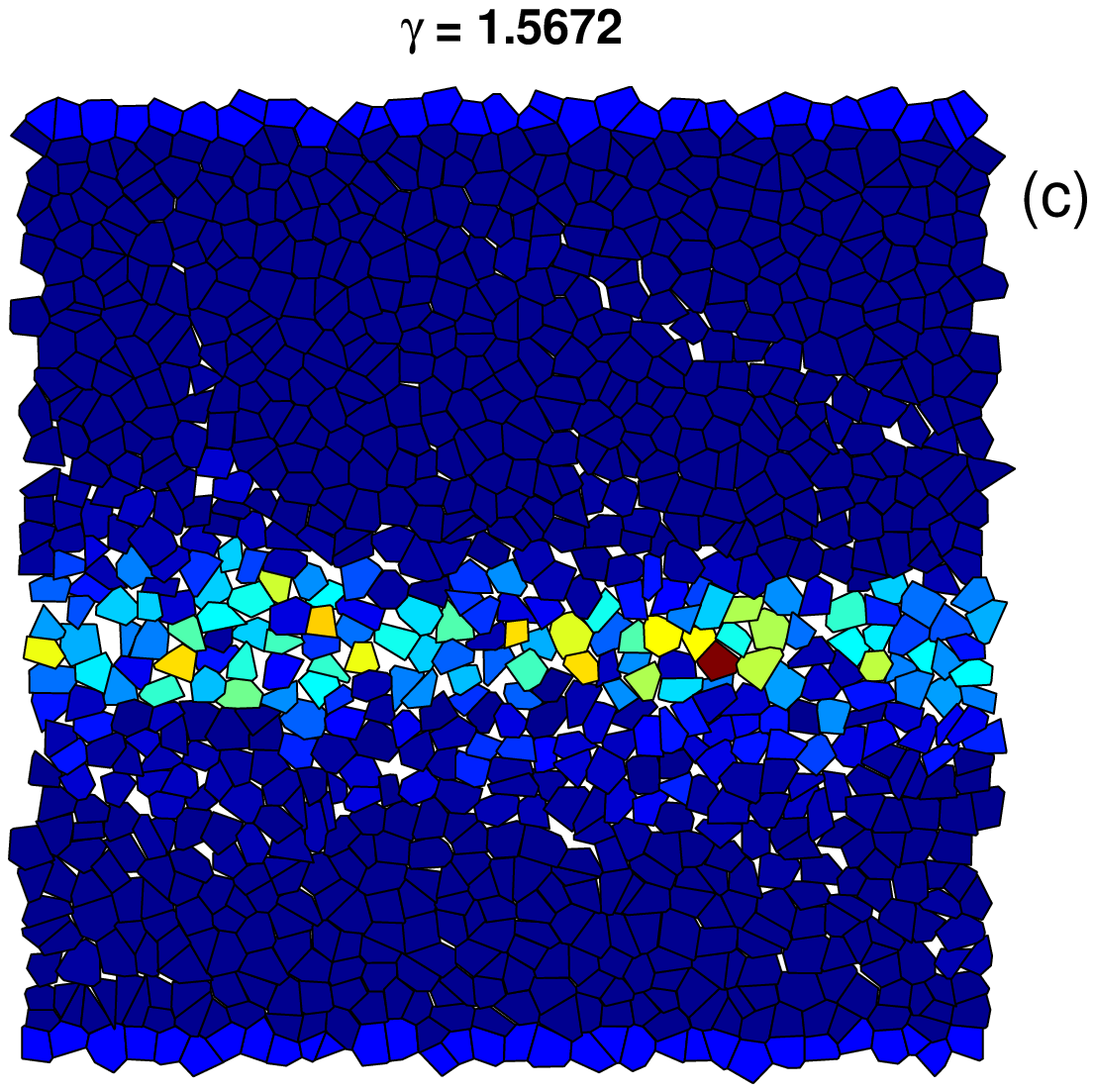}%
\includegraphics*[width=7.0cm]{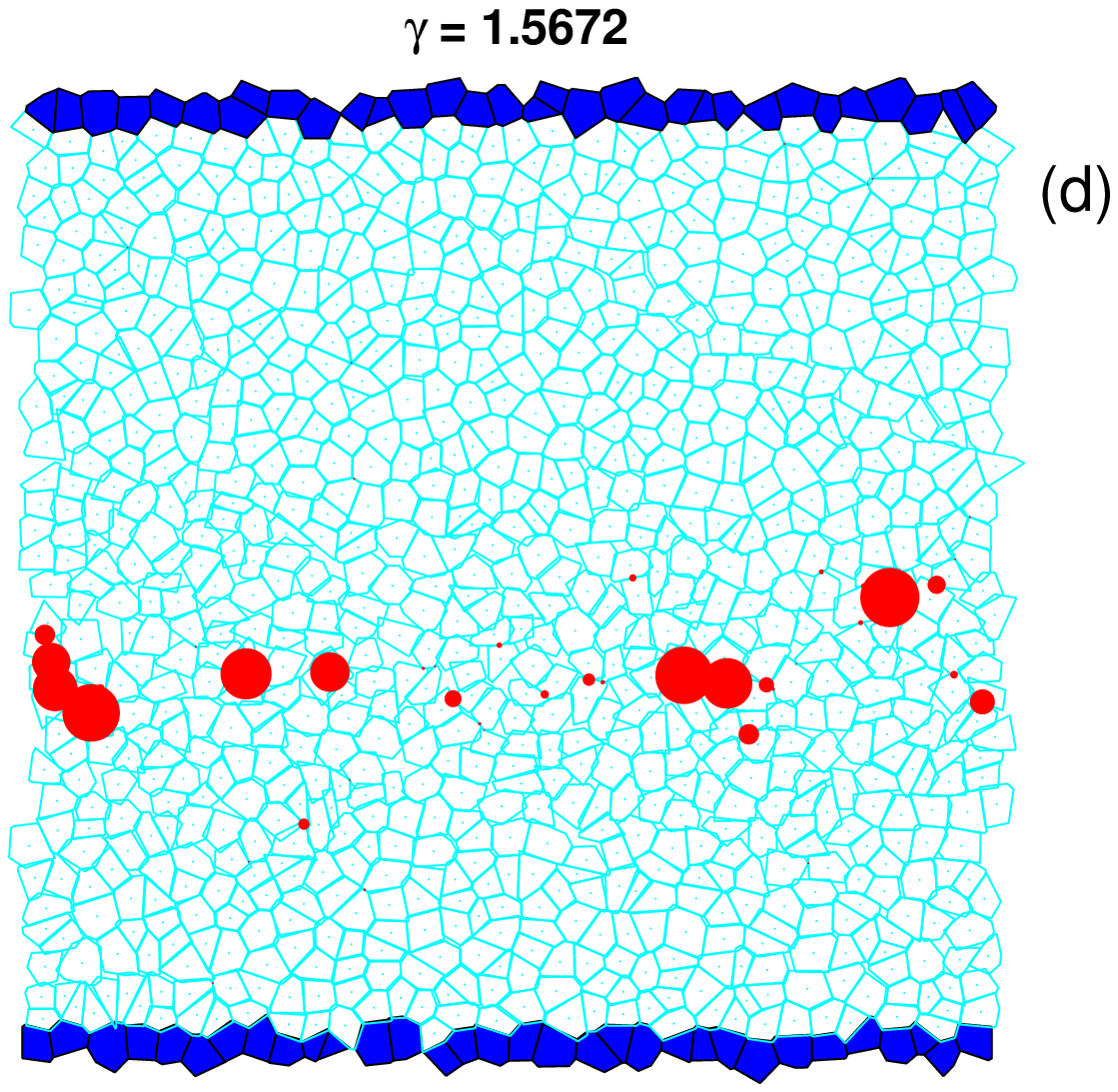}
\end{center}
\caption{\protect
         (Color online)
         Accumulation of elastic strain and overcome of the strength 
         $F_t/F_n$ of the material prior to the occurrence of an 
         avalanche. 
         In (a) the 
         kinetic energy of the system and (b) the ratio $F_s / F_n$ 
         showing the developed strength are shown, with red circles indicating 
         the strain value at which the snapshot in (c) and (d) are taken. 
         (c) The configuration and accumulated particle rotation just before 
         the avalanche. 
         (d) The elastic strain at the contacts 
         before the avalanche, where the diameter of the red dots is 
         proportional to the value of the deviatoric strain. 
         System size: $32\times32$ particles.}
\label{fig5}
\end{figure*}

In the case of infinitely rigid particles, subtracting the `stationary value' 
$K_0$ from the kinetic energy of the system, one would obtain zero 
between successive avalanches.
Our system is however composed of soft elastic particles,
and consequently a non-zero value is observed, as shown in Fig.~\ref{fig4}. 
This non-zero value is a numerical artifact \cite{pena07c} stemming from 
the calculation of the tangential contact forces, the soft elastic nature 
of the polygons, and the periodic boundary conditions that can trap some 
of the energy released during the avalanches. 
We checked that our present results are not affected by this numerical
noise and therefore we will not consider it in anymore detail.

The force needed to sustain the constant motion of the top and bottom layers 
can be measured in the simulation.  In the following, $F_s$ is the shear 
(horizontal) force applied to each wall, and $F_n$ is the normal force.
Figure~\ref{fig5} shows the occurrence of one avalanches and the associated 
strain accumulation for a system with $32 \times 32$ particles. 
We can see that the abrupt increment of kinetic energy of the system 
(Fig.~\ref{fig5}a), matches with the fall-off of the strength of the 
material $F_s / F_n$.

Figures \ref{fig5}c and \ref{fig5}d illustrate two different representations
of the same sample snapshot, immediately before the a\-va\-lanche.
Figure \ref{fig5}c shows the sample configuration and the 
rotation that the particles undergo for a shear band located at the center 
of the sample.
The colors of the particles are given by their accumulated rotation: 
the lighter the color the bigger the accumulated rotation. 
Figure~\ref{fig5}d shows the elastic strain at the contacts, which are
represented by red dots with a diameter proportional to its strain value.
Here, one can see that there is a strong localization of elastic strain 
along the shear band. This strain localization weakens the system and 
drives it to failure, since it promotes the occurrence of the
Coulomb limit condition related to the number of sliding contacts 
(see above).
In other words, the weakening of the system is due to both the
strain localization and the increase of the ratio of sliding contacts.

During the avalanche the system suffers a complete 
rearrangement in which the old sliding contacts are removed from the 
sliding condition and new contacts are generated. 
This rearrangement marks the beginning of a new stage of elastic strain 
accumulation that drives the system to the next avalanche.
\begin{figure}[!t]
\begin{center}
\includegraphics*[width=8.5cm]{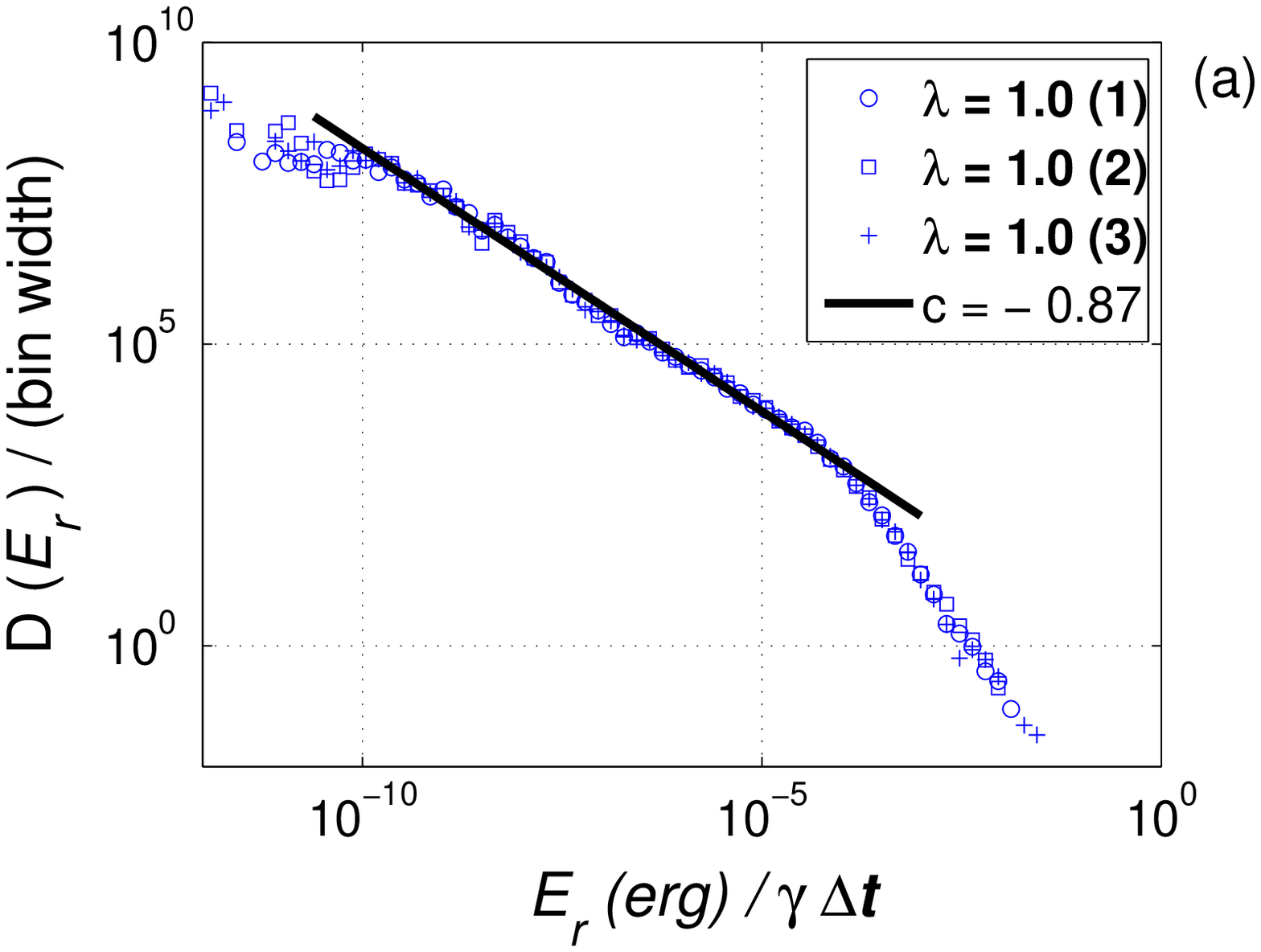}
\includegraphics*[width=8.5cm]{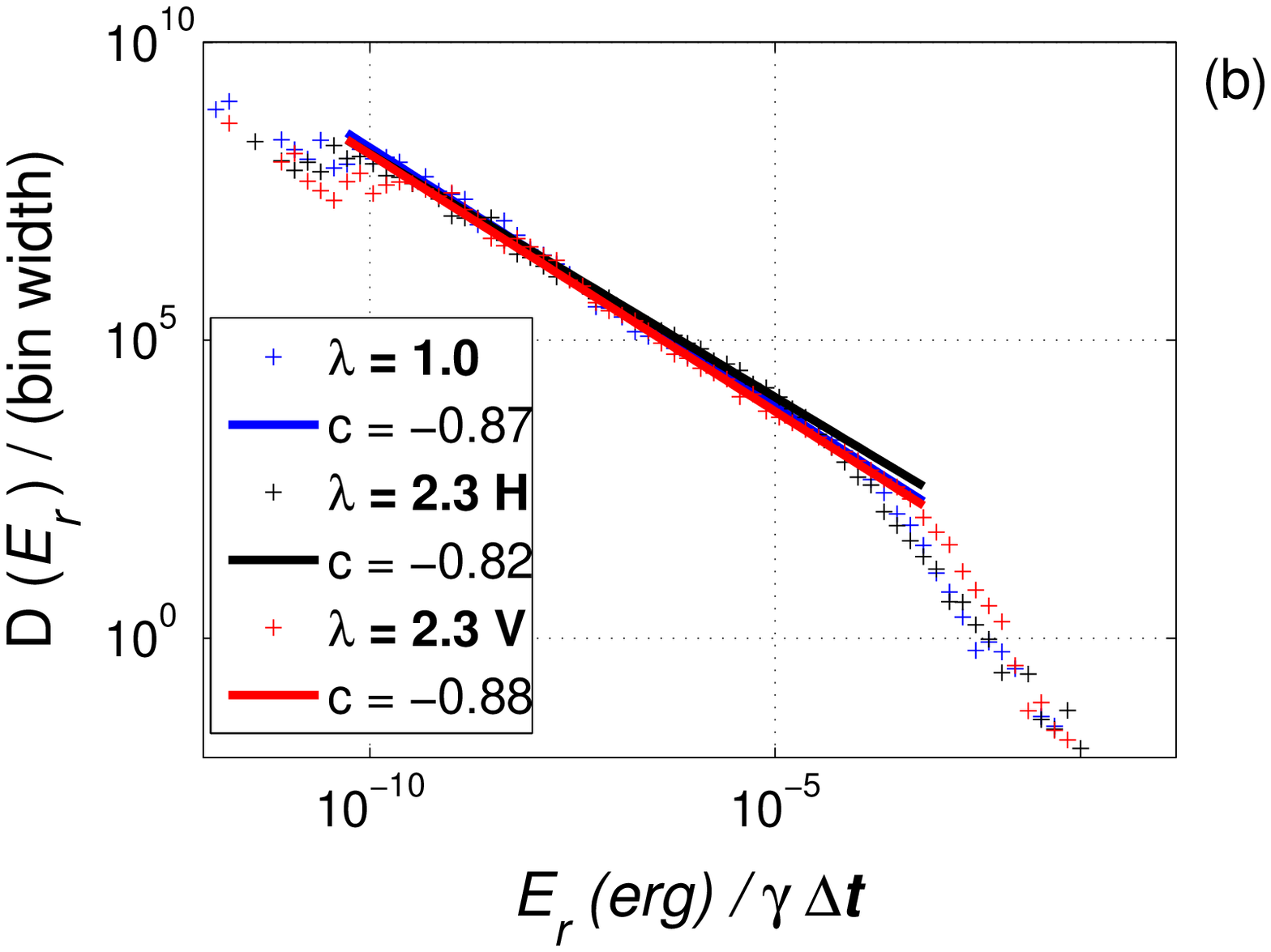}
\end{center}
\caption{\protect 
         (Color online)
         Log-log plot of the number of avalanches versus their 
         released energy $E_r$ for
         {\bf (a)} different configurations of isotropic particles and 
         {\bf (b)} for different 
         $\lambda$ values. Here 
         $\mu = 0.5$ and the system has $16\times16$ particles. 
         Logarithmic binning is used.}
\label{fig6}
\end{figure}
\begin{figure}[t]
\begin{center}
\includegraphics*[width=8.5cm]{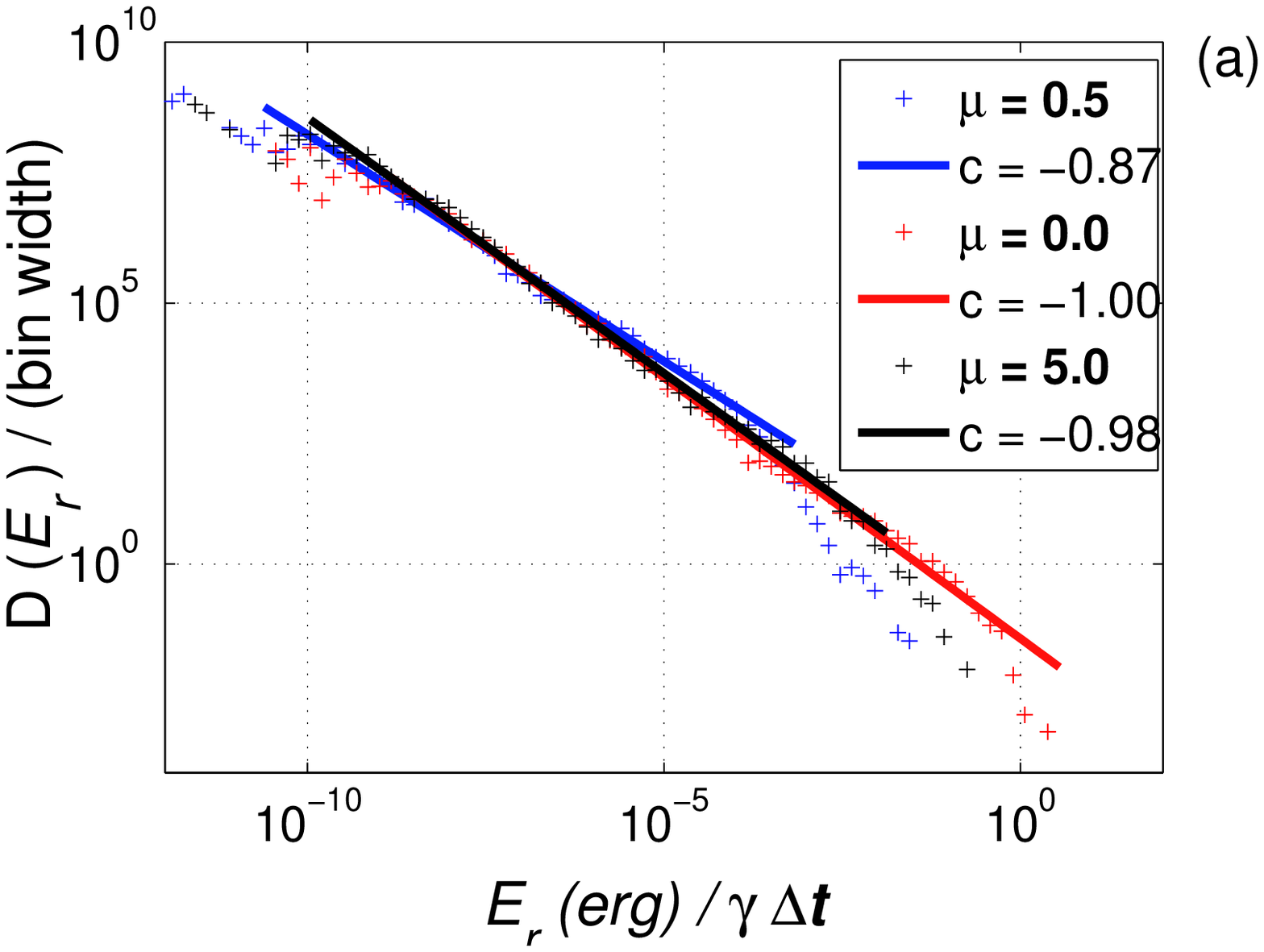}
\includegraphics*[width=8.5cm]{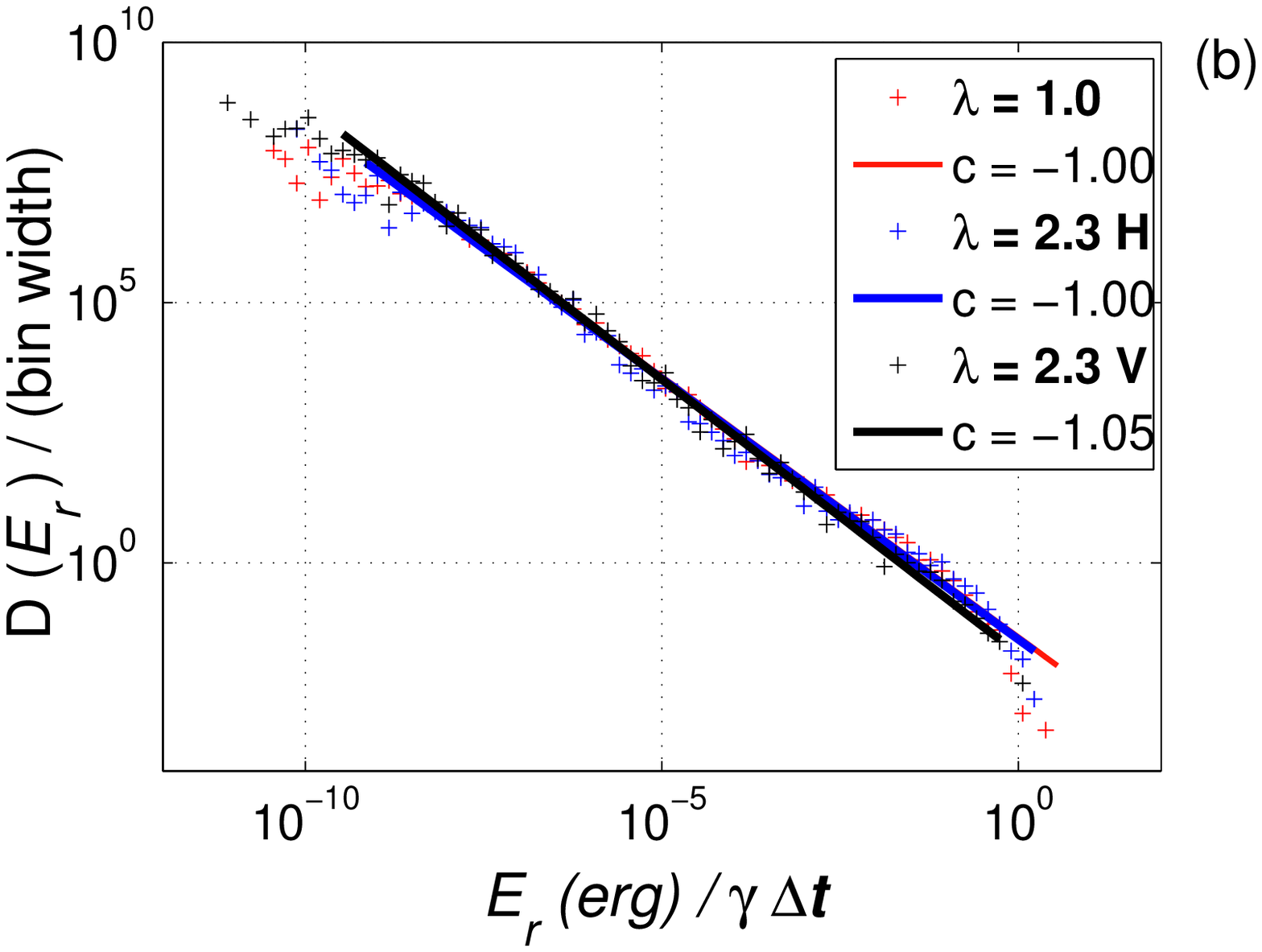}
\end{center}
\caption{\protect
         (Color online)
         The distribution $D(E_r)$ of the released energy $E_r$
         when
         {\bf (a)} varying interparticle friction coefficient $\mu$ with
         fixed $\lambda=1$ and 
         {\bf (b)} when varying $\lambda$ with fixed $\mu=0$. 
         The system has $16\times16$ particles and logarithmic binning 
         is used.}
\label{fig7}
\end{figure}

\section{The Gutenberg-Richter law in anisotropic granular media}
\label{sec:gr}

The distribution of earthquake magnitude is 
described by the Gutenberg-Richter law \cite{gutenberg54}. This law 
states that the number $n$ of earthquakes of magnitude $M$ is proportional 
to $n\sim 10^{-bM}$. Typically, the value of $b$ is equal to $1$ at most places, 
but may vary between $0.8$ and $1.5$ \cite{bikas97}. 
As we will see, 
the exponent $b$ will be an invariant property describing the occurrence of 
avalanches associated with sudden rearrangements of 
granular media under very slow shear.

To this end, we study the possible influence of the formation and 
evolution of the shear band on the distribution $D(E_r)$ of the 
released energy $E_r$ during an earthquake.
Since the magnitude of an earthquake is defined as the logarithm of the
released energy apart proper constants one finds 
$n\sim E_r^{-c}$, with the exponent $c$ varying typically in the range 
$0.8 < c < 1.1$~\cite{bikas97}.

In Figure~\ref{fig6}a we show the distribution $D(E_r)$ 
for three different initial configurations of isotropic samples,
corresponding to different seeds for the Voronoi 
Tessellation. 
All the distributions collapse and show a power law behavior 
over almost six orders of magnitude with an exponent of $c=0.87$ for the 
fitted straight line, which is within the observed range of values 
of the Gutenberg-Richter law.

In Fig.~\ref{fig6}b, the distributions for both isotropic and a\-niso\-tropic 
particles are shown.
Similarly, for all samples, the data sets are well fitted by a power law 
with an exponent $c$ ranging from $0.82$ to $0.89$, indicating a weak
influence of the particle shape on the distribution of the released energy.
The power law holds over six orders of magnitude.

Similar exponents ($0.80<c<0.95$) are obtained for other system sizes
in both isotropic and anisotropic cases and for the case when one 
considers 
the distribution for individual particles. 
From such results, one concludes that independent of the anisotropy 
there is a scale invariance of the system response according to the 
Gutenberg-Richter law.

We also study the influence of the friction coefficient $\mu$. 
In Fig.~\ref{fig7}a the distributions for the isotropic samples 
with different friction coefficients are plotted. 
The effect of friction in both cases is to slightly increase the exponent 
$c$, which holds for nearly seven orders of magnitude. 
The distributions for isotropic and aniso\-tropic samples with
$\mu= 0$ are presented in Fig.~\ref{fig7}b where 
no influence of particle shape is observed.

\section{Waiting times and Omori's law}
\label{sec:waiting}

Earthquakes usually occur as part of a sequence of events, in which 
the largest event is called the mainshock and the events prior and after 
the mainshock are foreshocks and aftershocks respectively \cite{scholz02}. 
The empirical law that describes the behavior of the temporal sequence of
avalanches is called Omori's law, and states that the number $n(t)$ 
of aftershocks decreases with the inverse of the time interval $t$
spanned from the last mainshock as
\begin{equation}
n(t) = \frac{d}{(1 + t)^p}
\label{eq:omori}
\end{equation}
where $d$ is an empirical constant and $p$ varies in the
range $0.7<p<1.5$ \cite{scholz02} with the most typical values
around one.

Before performing the calculation of waiting times of aftershocks in the 
system evolution, we have also to precisely define `mainshock'. 
Our definition is based on empirical observations \cite{scholz02}.
A new event is considered mainshock only when its released energy is larger 
than $1/10$ of the released energy of the last mainshock. 
When this happens the sequence of the aftershocks from the previous mainshock
is considered to be finished and a new sequence is calculated.
\begin{figure} [!t]
\begin{center}
\includegraphics*[width=9.0cm]{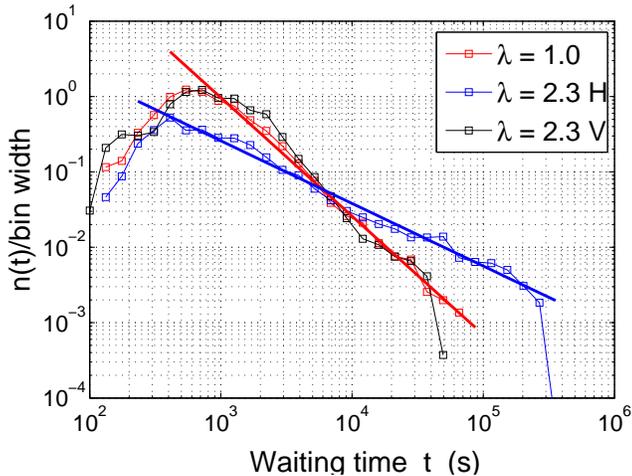}
\end{center}
\caption{\protect
         (Color online)
         Distribution $n(t)$ of waiting times for the sequence 
         of aftershocks in the numerical simulation. Isotropic 
         sample $\lambda = 1$ and anisotropic samples 
         $\lambda = 2.3$ are presented. The system has $16\times 16$ 
         particles.}
\label{fig8}
\end{figure}
\begin{figure*}[t]
\begin{center}
\includegraphics*[width=12.0cm]{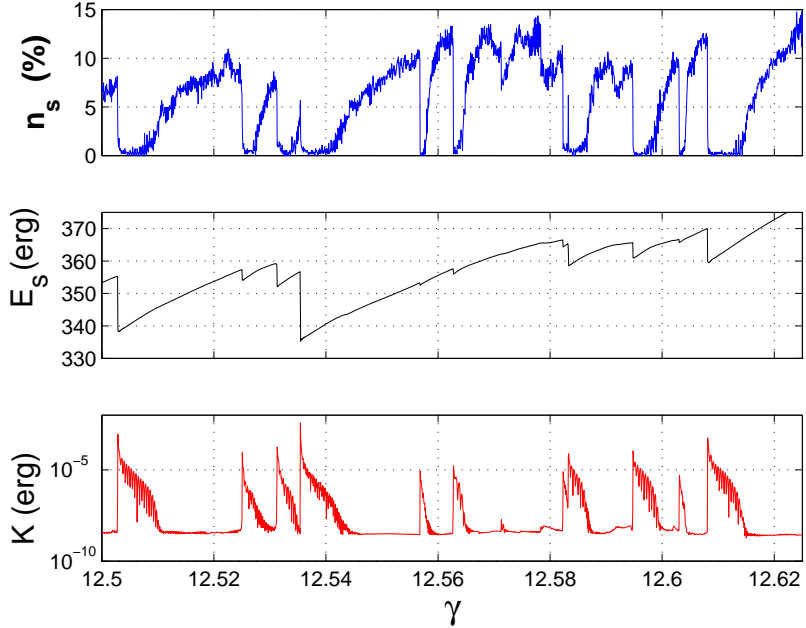}
\end{center}
\caption{\protect
         (Color online)
         Evolution of the relative number $n_s$ of sliding contacts, 
         the energy $E_s$ stored at the contacts and the total kinetic 
         energy $E_r$ as a function of the
         shear strain $\gamma$, for an isotropic sample ($\lambda=1$).}
\label{fig9}
\end{figure*}

In Fig.~\ref{fig8} the distribution of waiting times for isotropic and 
anisotropic systems 
are shown. 
Over more than three orders of magnitude
all the numerical results can be fitted using the expression in 
Eq.~(\ref{eq:omori}),
with exponents $p=1.57$ for $\lambda = 1$, $p=1.61$ for $\lambda = 2.3$ V, 
and $p=0.83$ for $\lambda = 2.3$ H. 
While for $\lambda = 2.3$ H one observes an exponent
within the typical range found in fault gouges, for $\lambda = 1$ and 
$\lambda = 2.3$ V one finds a clear deviation from the observed values.
This indicates not only that anisotropy should be ubiquitous 
in fault gouges, but also that the most stable configurations are the 
most common, as explained below.

Using this influence of the initial configuration of aniso\-tropic samples 
on the stability of the system, we will next explain how to detect at the 
macro-mechanical level the presence of anisotropic particles within the gouge.

The anisotropic sample $\lambda = 2.3$ H with particles oriented parallel to 
the shear direction exhibits a more stable configuration, than the other 
two cases $\lambda=1$ and $\lambda=2.3$ V. In this sample, 
the induced torque on the particles is minimized
and the main deformation modes, sliding and rolling of the particles, are
highly suppressed by the fixed boundary conditions allowing no dilation in 
vertical direction. The hindrance of the deformation modes produces a 
larger temporal stability and also a larger mechanical stability. The 
larger temporal stability makes the occurrence of the events less 
frequent in time, i.e.~a slower decay of the waiting times. The larger 
mechanical stability results in a smaller probability of failure for a 
given value of stiffness as discussed in Sec.~\ref{sec:precursors} below.  
On the contrary, the configuration of anisotropic samples $\lambda = 2.3$ V, 
with particles oriented perpendicular to the shear direction, maximizes 
the induced torque on the particles and results in a less stable 
configuration.  This configuration yields smaller temporal and mechanical 
stability. The smaller temporal stability is observed in the decay of the 
waiting times, that is slightly faster than the one of the isotropic sample. 
The smaller mechanical stability of sample $\lambda = 2.3$ V is manifested 
in the larger probability of failure for a given value of stiffness 
compared to the other samples. 

Therefore, by looking at the decay rate of the aftershock sequences one might
be able to explain the variation of the exponent $p$ in realistic 
earthquake sequences, by the existence of 
anisotropic gouge in the fault zone.
It is important to say that for a more realistic representation of the 
earthquake process the crushing of particles should also be taken into 
account. 
Nevertheless, the absence of particle crushing is a suitable approximation 
for the case of young fault gouges.

\section{Weakening and mechanical stability of the system}
\label{sec:precursors}

In this section we study the relationship between the occurrence of 
avalanches and the weakening of the system. The weakening process results 
from the release of energy due to previous accumulation of strain at the 
contact level and contacts reaching the sliding condition.

In Fig.~\ref{fig9} we show the relative number of sliding contacts  
$n_s$,
the stored energy at the contacts $E_s$ and the total kinetic 
energy $E_r$. 
The stored elastic energy $E_s$ at the contacts is calculated 
as $E_s=1/2 \; (k_n \delta^2 + k_t \xi^2)$, where 
$\delta$ and  $\xi$ are the spring elongations in the normal and 
tangential directions respectively.
Figure \ref{fig9} shows that between avalanches the relative number 
of sliding contacts increases with the shear strain. It makes the system 
weaker and indicates that the system is constantly accumulating elastic 
energy $E_s$ at the contacts. 
The weakening of the system persists 
until failure, where the kinetic energy increases by several orders of 
magnitude. At this stage, the structure of the system is rearranged, the 
stored energy at the contacts is released and the contacts do not fulfill
the sliding condition. 
All the events in the kinetic energy are associated with both drops in 
the ratio $n_s$ and drops in the stored energy. 

At the macromechanical level the weakening of the system is observed by 
looking at the evolution of the shear stress with the shear strain.
After each stress drop the system experiences a rearrangement.
This new configuration produces a temporal stability, in which the strength 
builds up.  At this stage the system sticks and the accumulation of strain 
takes place. 
In this softening regime the 
system approaches failure and when the strength of the material is 
overcome, the system slips. 

The mechanical stability of the system is studied through the stiffness 
value $\Delta\tau/\Delta \gamma$ that the system presents at failure. 
To this end, we calculate the conditional probability 
$P(A_E|\Delta \tau / \Delta \gamma)$ for the occurrence of an avalanche 
event $A_E$ given a stiffness value $\Delta \tau / \Delta \gamma$.
Since data only provides the complementary conditional probability
$P(\Delta \tau / \Delta \gamma|A_E)$ we use the Bayes theorem \cite{papoulis02}, 
to obtain
\begin{equation}
P(A_E|\Delta \tau / \Delta \gamma) = \frac {P(\Delta \tau / 
      \Delta \gamma|A_E) P(A_E)}{P(\Delta \tau / \Delta \gamma)} .
\label{eq:Bayes}
\end{equation}

Having a certain time increment $d t$ in the data, the conditional 
probability  $P(\Delta \tau / \Delta \gamma|A_E)$ gives the
probability of observing a stiffness value $\Delta \tau / \Delta \gamma$ 
at $t-d t$ when at time $t$ an avalanche (symbolized by $A_E$)
occurs.
The conditional probability takes into account all the avalanches $A_E$ 
within a total time interval of $5\cdot10^6$ s. 
To this end, we discretize the total time interval in time increments 
of $d t = \Delta \gamma / \dot{\gamma}$. 
Since the shear rate $\dot{\gamma} =  1.25\cdot10^{-5}$ s
$^{-1}$ and $\Delta \gamma = 0.0016$, the time increment 
is $d t = 128$ s is also constant in our simulation. 
We select the same $dt$ for both isotropic 
and anisotropic systems. 
The probability $P(A_E)$ is fraction between the total number of 
avalanches and the total number of time increments, 
and similarly $P(\Delta \tau / \Delta \gamma)$ is the probability 
distribution of $\Delta \tau / \Delta \gamma$. 
\begin{figure} [!t]
\begin{center}
\includegraphics*[width=8.5cm]{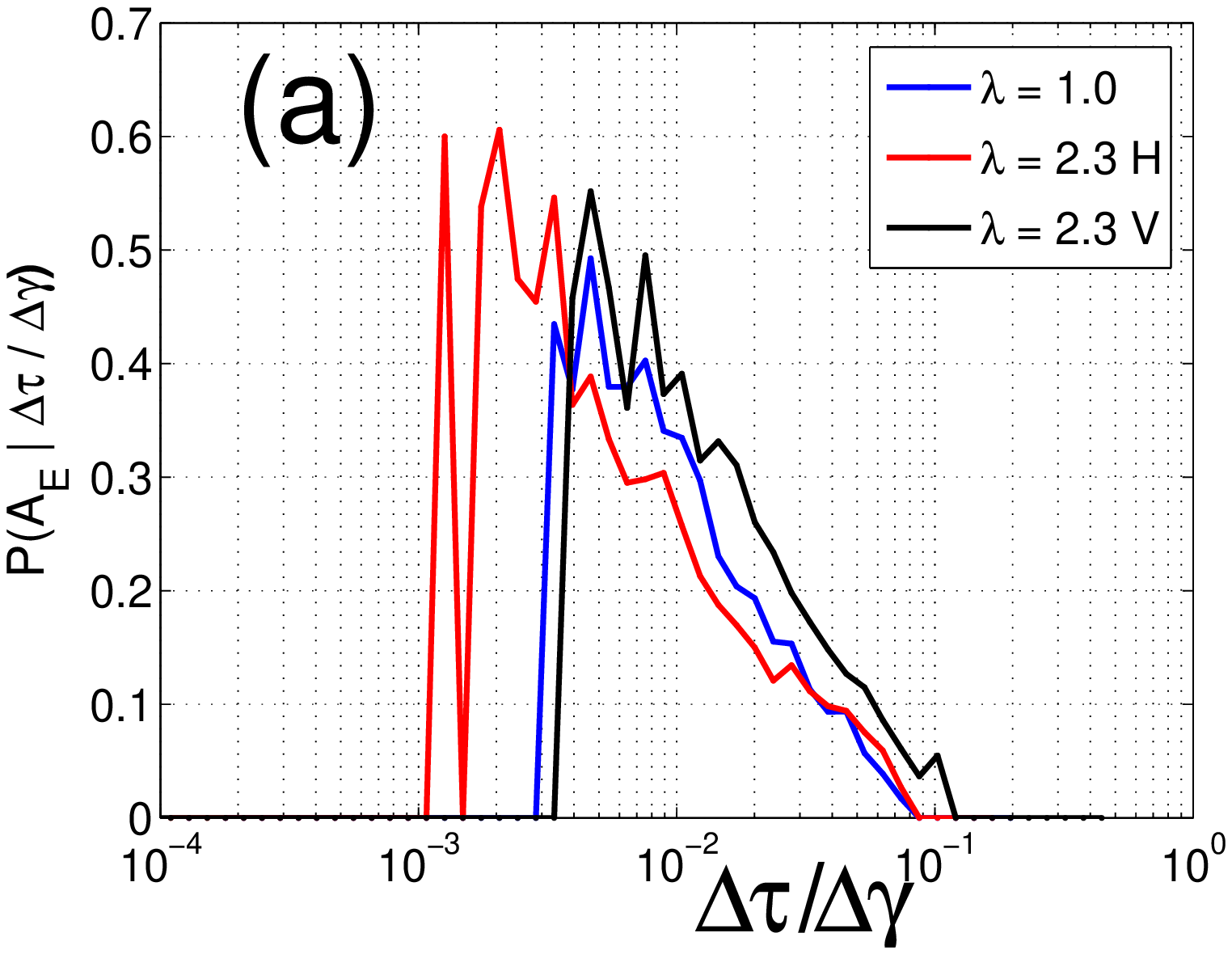}
\includegraphics*[width=8.5cm]{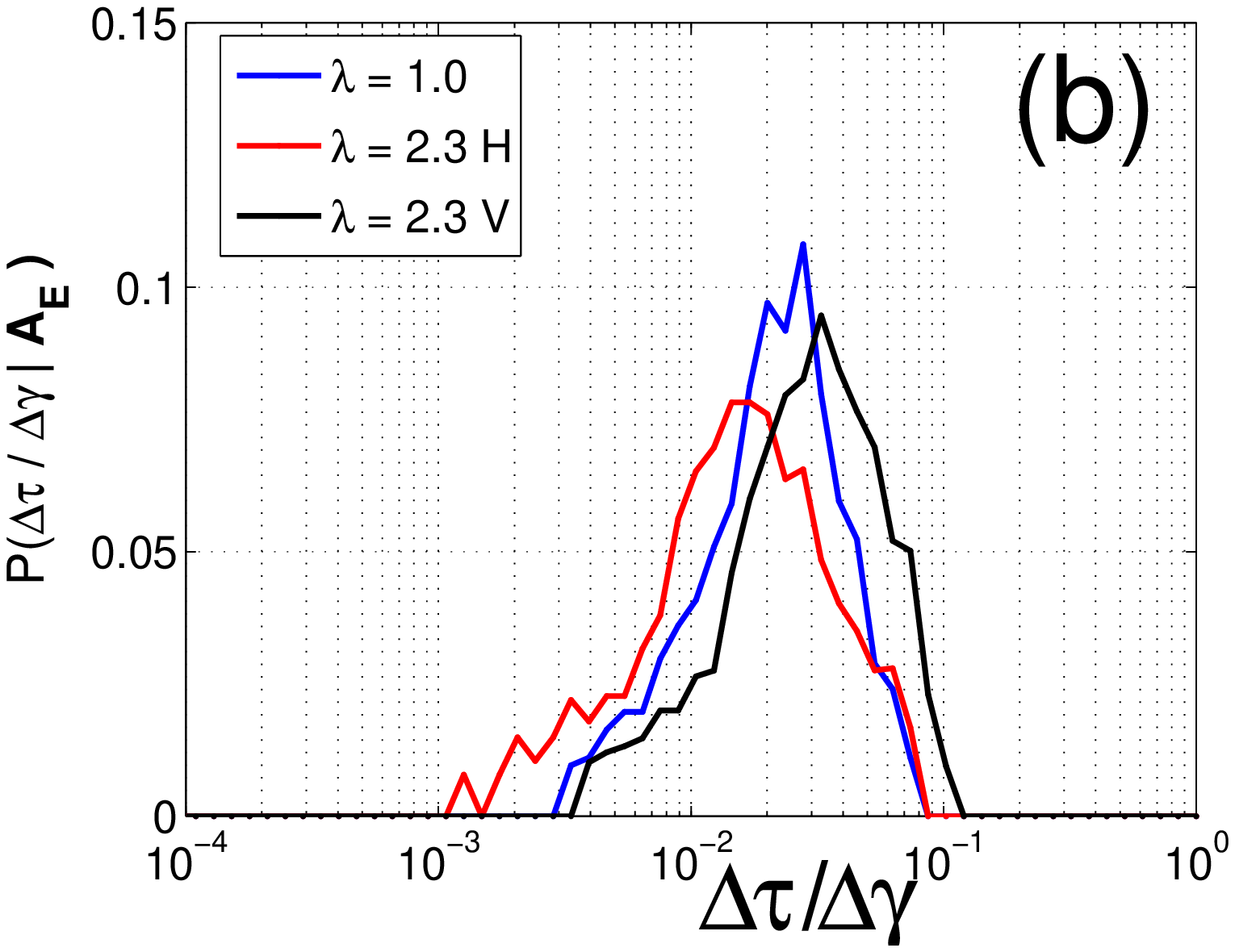}
\end{center}
\caption{\protect
         (Color online)
         Conditional probability distributions (a) $P(A_E|\Delta 
         \tau / \Delta \gamma)$ of the occurrence of an avalanche 
         $A_E$ given a stiffness $\Delta \tau / \Delta \gamma$ and 
         (b) $P(\Delta \tau / \Delta \gamma|A_E)$ of having a 
         stiffness $\Delta \tau / \Delta \gamma$ at the occurrence 
         of an avalanche. System size $16 \times 16$. Isotropic 
         $\lambda = 1$ and anisotropic samples $\lambda = 2.3$ 
         are presented. Labels $H$ and $V$ correspond to horizontal 
         and vertical samples.}
\label{fig10}
\end{figure}
In Fig.~\ref{fig10}, the conditional probabilities 
$P(A_E|\Delta \tau / \Delta \gamma)$ and  
$P(\Delta \tau / \Delta \gamma|A_E)$ for both the isotropic system 
$\lambda = 1$ and the anisotropic system $\lambda = 2.3$ are shown. 
For all the samples, the conditional probability 
$P(A_E|\Delta \tau / \Delta \gamma)$ decreases logarithmically
with the stiffness: the stiffer the system the smaller is the 
probability of failure, as shown in Fig~\ref{fig10}a, yielding
\begin{equation}
P(A_E|\Delta \tau / \Delta \gamma) \sim q \log{\left (
        \Delta \tau / \Delta \gamma
        \right )} .
\end{equation}
Further, from Figs.~\ref{fig10}a and \ref{fig10}b,
anisotropic samples compared to the isotropic ones explore 
a different range of stiffness at failure due to the larger rotational 
frustration that the elongated particles undergo.
This is specially true for particles oriented along the shear direction
($\lambda = 2.3$ H), due to its more stable structure. 
The exponents of the tail of the distributions are $q=-42.1$ 
for $\lambda = 1$, $q=-28.6$ for $\lambda = 2.3$ V 
and $q=-30.78$ for $\lambda = 2.3$ H (see Fig.~\ref{fig10}a). 
\begin{figure} [!t]
\begin{center}
\includegraphics*[width=8.5cm]{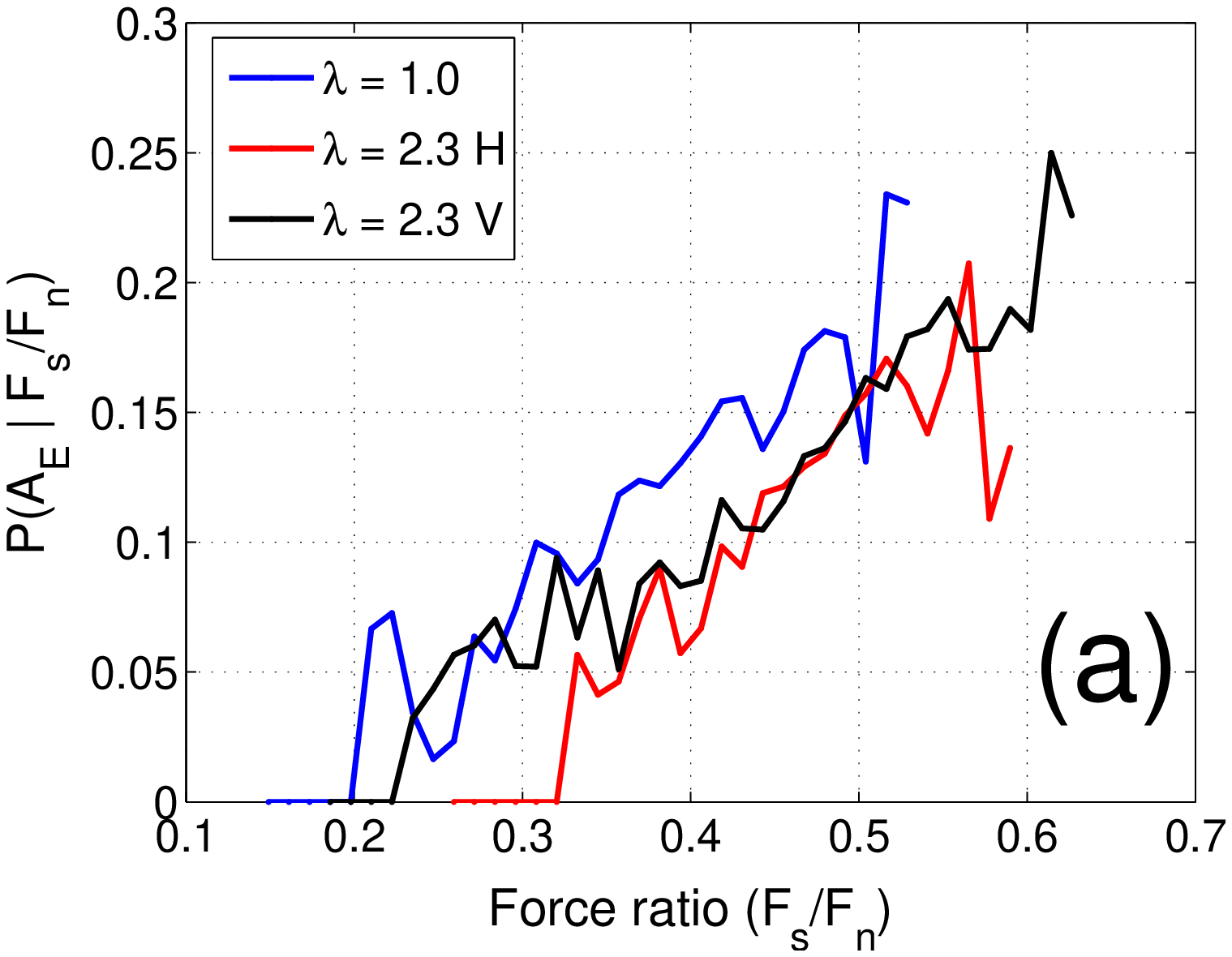}
\includegraphics*[width=8.5cm]{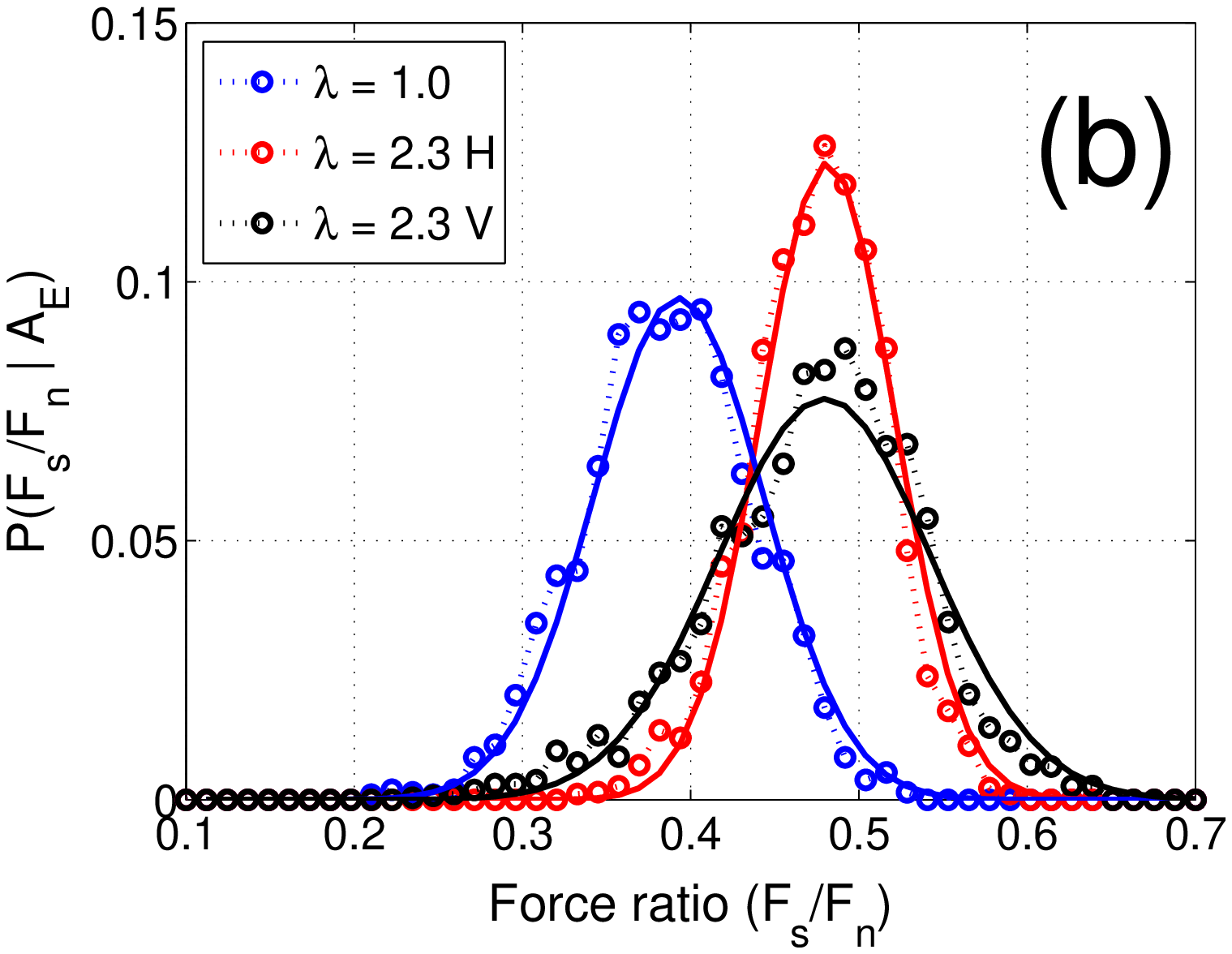}
\end{center}
\caption{\protect
         (Color online)
         Conditional probability distributions (a) $P(A_E|F_s/F_n)$ 
         of the occurrence of an avalanche $A_E$ given a frictional 
         strength $F_s/F_n$ and (b) $P(F_s/F_n|A_E)$ of having a 
         frictional strength $F_s/F_n$ at the occurrence of an 
         avalanche. System size $16 \times 16$. Isotropic $\lambda = 1$ 
         and anisotropic samples $\lambda = 2.3$ are presented. 
         Labels $H$ and $V$ correspond to horizontal and vertical 
         samples. The $P(F_s/F_n|A_E)$ follows a normal distribution, 
         except for sample $\lambda = 2.3$ V. Solid lines represent 
         the normal distribution of the data.}
\label{fig11}
\end{figure}

The same analysis can be performed for the occurrence of an avalanche 
$A_E$ given a force ratio $F_s/F_n$ value. 
In Figure~\ref{fig11}, the conditional probabilities $P(A_E|F_s/F_n)$ and  
$P(F_s/F_n|A_E)$ for the same cases as in Fig.~\ref{fig10}.
As shown in Fig.~\ref{fig11}a, the conditional probabili\-ty \- 
$P(A_E \vert F_s/F_n)$ increases approximately linearly with $F_s/F_n$: 
The higher the mobilized strength $F_s/F_n$  the higher the probability 
of failure.
In general, the anisotropic samples are able to mobilize 
higher frictional strength than the isotropic sample and,
for the same force ratio, present a lower probability 
\- $P(A_E \vert$ $F_s / F_n)$. 
This is certainly due to the influence of particle 
shape a\-niso\-tropy on the global strength of the material \cite{pena07a}.

The probability distribution $P ( F_s / F_n \vert A_E)$ shown in 
\- Fig.~\ref{fig11}b of having a force ratio $F_s/F_n$ at the occurrence 
of an ava\-lan\-che $A_E$ follows typically a normal distribution, 
with a mean value larger in the anisotropic samples.
Considering only the anisotropic cases, while one observes the same mean, the
standard deviation is larger for the less stable configuration V, as expected.

\begin{figure} [!t]
\begin{center}
\includegraphics*[width=8.5cm]{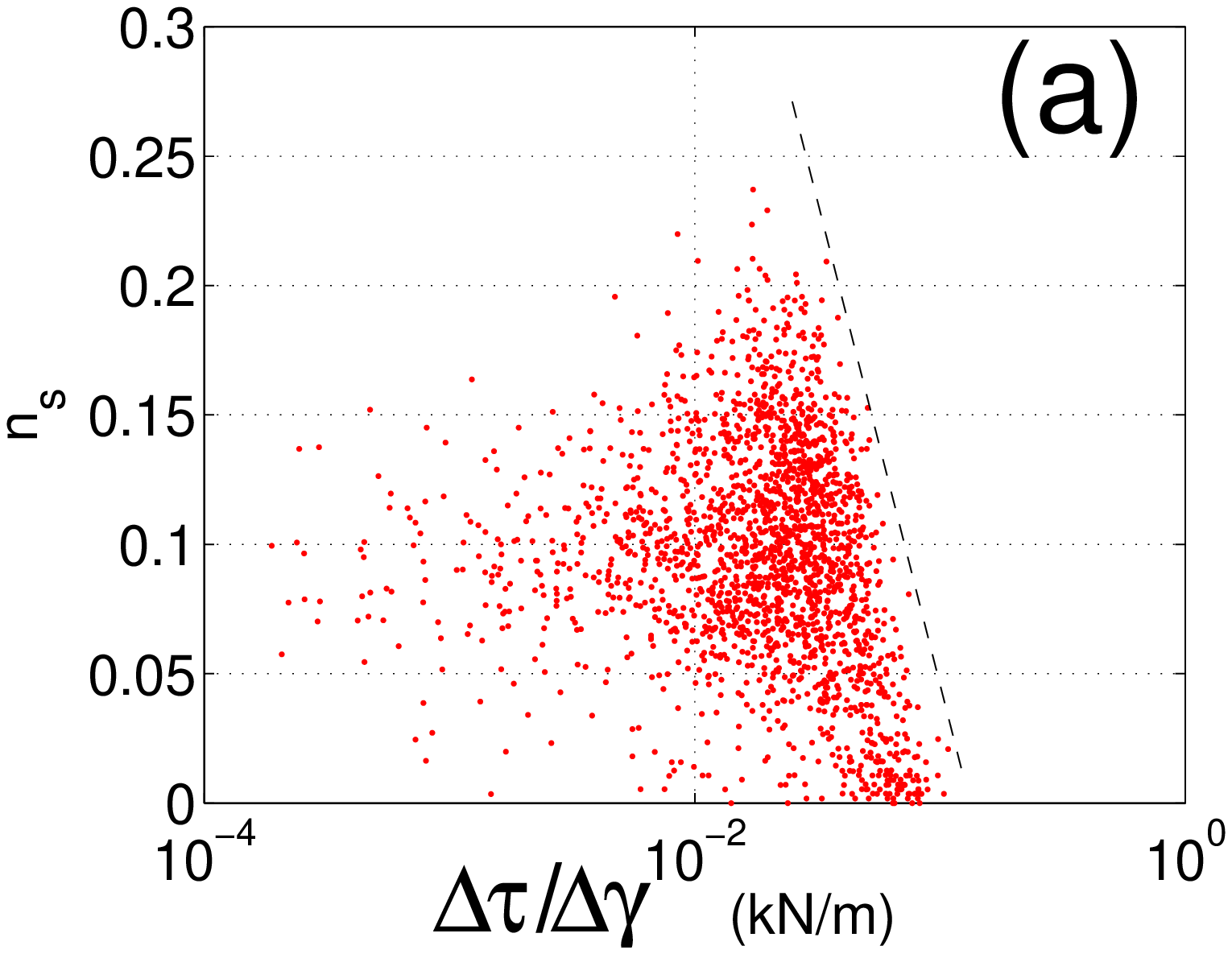}
\includegraphics*[width=8.5cm]{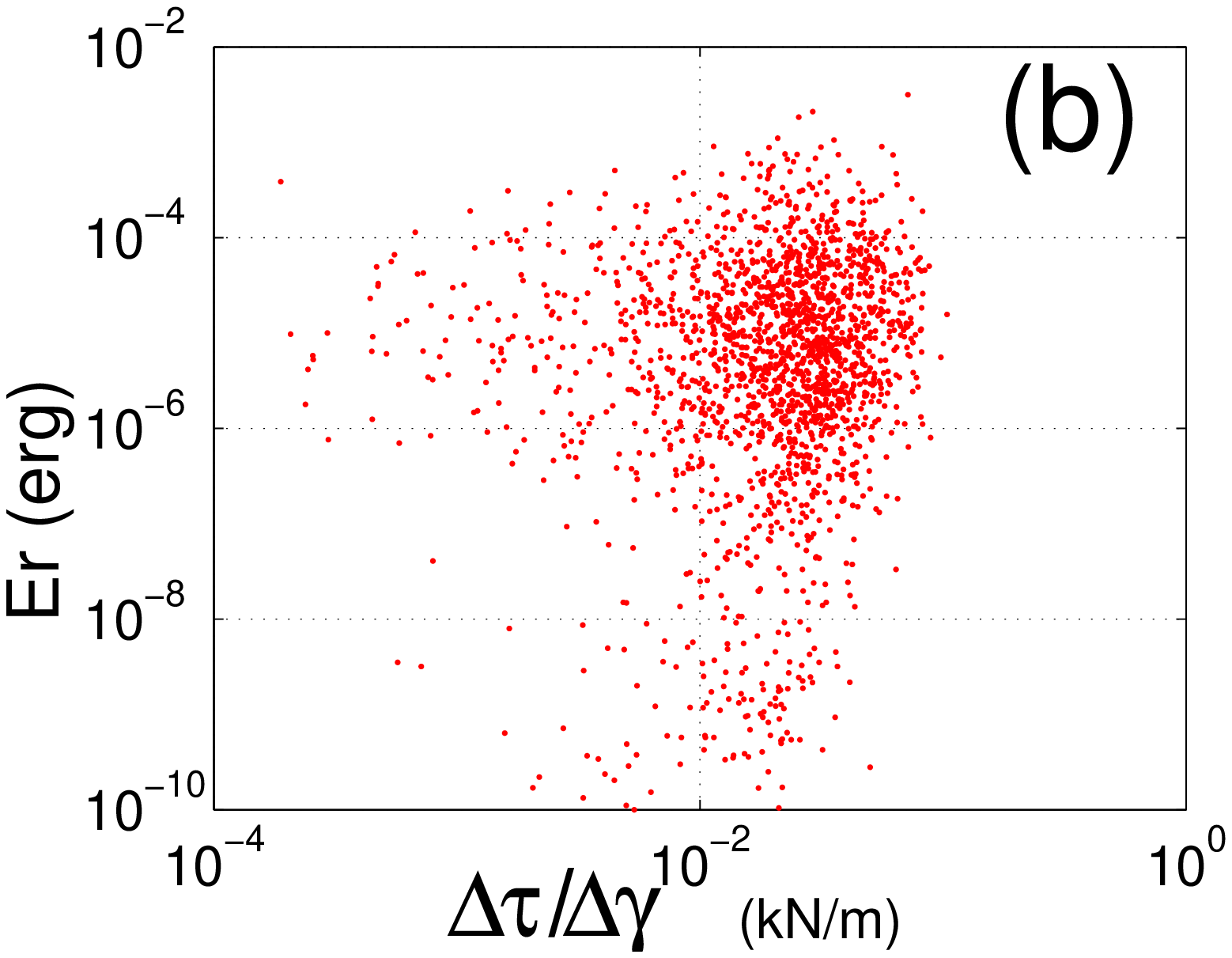}
\end{center}
\vspace{0.5cm}
\caption{\protect
         (Color online)
         Relationship between (a) sliding contact ratio $n_s$ vs. 
         stiffness at failure, (b) released energy $E_r$ of the avalanche vs 
         stiffness at failure. Data correspond to an isotropic 
         sample with size $16 \times 16$ particles.}
\label{fig12}
\end{figure}


Finally, in Figure~\ref{fig12} the sliding contact ratio $n_s$ and the 
stiffness at failure of the system are presented. The stiffness is also 
compared to the released energy of the avalanches. Although no clear correlation 
between the parameters can be observed, the maximum value of stiffness 
that the system can present is bounded by $n_s$, as illustrated in 
Fig.~\ref{fig12}a.
Larger $n_s$ implies smaller stiffness. 
In Fig.~\ref{fig12}b one sees a weak correlation between the stiffness and 
the released energy, with a slight tendency of $E_r$ increasing with
the stiffness. 
From these observations, it is clear that the accumulation of strain at the 
contacts is not the only important agent in the weakening process. 
Therefore, additional ingredients have to be 
taken into account for a more exhaustive analysis. 

\section{Concluding remarks}
\label{sec:conclusions}

In this paper we highlighted the importance of particle shape on the 
occurrence of avalanches in granular systems.
To this end, we used shear cells with periodic boundary conditions to 
mimic the behavior of tectonic faults with fixed boundaries. 

We found that the dynamics of the granular system is characterized by 
discrete avalanches spanning several order of magnitude similar to  
crackling noise \cite{sykes99,sethna01}. 
We calculated the probability distribution of the energy released in 
avalanches, and found it to be 
in very good agreement with the Gutenberg-Richter law 
for samples with different particle anisotropy and different system sizes. 
Consequently the exponent of the released energy distribution can be
seen as an invariant property of such systems.

We also studied the temporal distribution of event sequences after a 
mainshock. We found that the number of aftershocks decreases with a power
of the inverse of time.
We could fit the sequences of waiting times 
of the aftershocks with the empirical expression and obtained exponents 
in the range $0.7 < p < 1.6$, similarly to what is observed in real
observations according to Omoris law.
The anisotropic sample H 
exhibits a larger temporal stability 
making the temporal occurrence of the avalanches sparser, 
due to its larger frustration of rotation in the corresponding
initial configuration.

The larger temporal stability observed at the ma\-cro-\-me\-cha\-nical level 
can be therefore taken as an indication of the existence of anisotropic 
material within the shear zone. 
This could potentially explain the variation of the exponent $p$ observed 
in realistic earthquake sequences.

The dynamics of the system was also related to the stick-slip 
process \cite{nasuno98,feder91}. 
When one avalanche begins the system slips, while between two 
successive avalanches, the system sticks, accumulates 
elastic energy, and becomes weaker because of the increase of the 
sliding contacts $n_s$. 
We characterized the weakening of the system by the 
stiffness $\Delta \tau / \Delta \gamma$ and derived the conditional 
probability $P(A_E|\Delta \tau / \Delta \gamma)$ for an avalanche to 
occur given a stiffness value $\Delta \tau / \Delta \gamma$. 
We found that $P(A_E|\Delta \tau / \Delta \gamma)$ decreases logarithmically
with the stiffness and with a decay rate larger for isotropic samples than for
anisotropic ones.
Concerning frictional strength we found that the probability of an 
ava\-lanche to occur increases with the force ratio $F_s/F_n$. 

The results concerning the conditional probabilities uncovered
that anisotropic samples can explore a wider range of stiffnesses and
force ratios than the isotropic sample. 
This is due to the larger kinematic constraint that anisotropic particles 
undergo during shear.
Further, in general, since the initial configuration corresponding to 
sample H is the most stable configuration with respect to shearing,
it is the one with lower probability of failure. 

In previous works concerning the avalanches in granular 
piles \cite{staron05,staron02}, some avalanche precursors associated to 
the onset of fluidized regions of sliding contacts were found. These 
fluidized regions were located in the weak network of contacts. 
This weak network comprises contacts transmitting forces smaller 
than the average force and therefore has a minimal contribution to 
the stress state \cite{radjai96b,radjai98b}. 
It is known that \cite{staron05,radjai98b,majmudar05}, while the strong 
contact network 
is responsible for the 
strength and stability of the packing, the weak contact network plays an 
important role in the destabilization process. 
In this scope, since precursors are expected to be related to sharp changes 
in the micro-structure of the granular packing \cite{staron06}, it would
be important to study also the network of contact forces as a function of
the anisotropy, in particular as a function of the major axis orientation.

Finally, 
although the results from our numerical model show good agreement with the 
processes observed in nature, we are aware of the challenges to have a more 
realistic simulation of fault zones \cite{alonso07}. 
In future work one should also extend this approach to the three-dimensional
case, consider the development of transparent (or 
absorbing) boundary condition, since the acoustic emission after an 
avalanche are trapped due to the periodic boundary conditions, and also 
implement the grain fragmentation, since natural earthquakes result 
from the combined effect of frictional instabilities and rock fragmentation.

\begin{acknowledgements}
The authors thank L.~Arcangelis for useful discussions and
also the support by German-Israeli Foundation and
by {\it Deutsche Forschungsgemeinschaft}, under the project
HE\-273\-2781.
PGL thanks support by {\it Deutsche Forschungsgemeinschaft}, under
the project LI 1599/1-1. HJH thanks the Max Planck prize.
\end{acknowledgements}



\end{document}